\documentclass[sigconf,authorversion]{acmart}
\AtBeginDocument{%
  \providecommand\BibTeX{{%
    \normalfont B\kern-0.5em{\scshape i\kern-0.25em b}\kern-0.8em\TeX}}}

\setcopyright{acmcopyright}
\copyrightyear{2021}
\acmYear{2021}
\setcopyright{acmlicensed}\acmConference[MuC '21]{Mensch und Computer 2021}{September 5--8, 2021}{Ingolstadt, Germany}
\acmBooktitle{Mensch und Computer 2021 (MuC '21), September 5--8, 2021, Ingolstadt, Germany}
\acmPrice{15.00}
\acmDOI{10.1145/3473856.3474024}
\acmISBN{978-1-4503-8645-6/21/09}

\begin{document}

\title[Investigating the Sense of Presence Between Handcrafted and Panorama Environments]{Investigating the Sense of Presence Between Handcrafted and Panorama Based Virtual Environments}


\author{Alexander Schäfer}

\orcid{0000-0003-2356-5285}
\affiliation{%
  \institution{TU Kaiserslautern}
  \streetaddress{P.O. Box 1212}
  \city{Kaiserslautern}
  \state{Rheinland Pfalz}
  \country{Germany}
  \postcode{66953}
}
\email{alexander.schaefer@dfki.de}

\author{Gerd Reis}
\affiliation{%
  \institution{German Research Center for Artificial Intelligence}
  \streetaddress{Trippstadterstraße 122}
  \city{Kaiserslautern}
  \country{Germany}
}
\email{gerd.reis@dfki.de}
\author{Didier Stricker}
\affiliation{%
  \institution{German Research Center for Artificial Intelligence, TU Kaiserslautern}
  \city{Kaiserslautern}
  \country{Germany}
}
\email{didier.stricker@dfki.de}

\renewcommand{\shortauthors}{Schäfer et al.}

\begin{abstract}
Virtual Reality applications are becoming increasingly  mature. The requirements and complexity of such systems is steadily increasing. Realistic and detailed environments are often omitted in order to concentrate on the interaction possibilities within the application. Creating an accurate and realistic virtual environment is not a task for laypeople, but for experts in 3D design and modeling. To save costs and avoid hiring experts, panorama images are often used to create realistic looking virtual environments. These images can be captured and provided by non-experts.
Panorama images are an alternative to handcrafted 3D models in many cases because they offer immersion and a scene can be captured in great detail with the touch of a button. This work investigates whether it is advisable to recreate an environment in detail by hand or whether it is recommended to use panorama images for virtual environments in certain scenarios. For this purpose, an interactive virtual environment was created in which a handmade 3D environment is almost indistinguishable from an environment created with panorama images. Interactive elements were added and a user study was conducted to investigate the effect of both environments to the user. The study conducted indicates that panorama images can be a useful substitute for 3D modeled environments.
\end{abstract}




\maketitle

\section{Introduction}
High-quality 3D assets for immersive virtual environments are expensive and usually not widely used in applications due to lack of money and time. Virtual replicas of buildings or rooms are even more expensive, as the fine details are often neglected by 3D designers due to time constraints in projects. Panorama images can be used to capture a scene with the touch of a button which can then be viewed in Virtual Reality (VR). Furthermore, a panorama-based environment is less computationally intensive compared to high visual quality 3D scenes. Using panorama images for VR and Mixed Reality (MR) is becoming increasingly popular. As an example, Schäfer et al. \cite{schafer2019towards} used panorama images to create a photorealistic VR meeting room. Sayyad et al. \cite{sayyad2017panotrace} implemented a system where panorama images are blended together with 3D objects via texture inpainting. Rhee et al. \cite{rhee2017mr360} implemented a system that seamlessly composites 3D virtual objects into a  360° panoramic video. The number of applications which use panorama images and videos is steadily increasing and new applications are constantly being developed.\par 
However, to the best of our knowledge, no study has been conducted which compares user perception between a panorama-based environment and a 3D modeled counterpart. This work aims to discover if it is worthwhile to create a virtual replica of an environment or if it is enough to capture the desired scene in panorama images and then present it to the user. For this purpose, a virtual room was recreated in great detail which is compared with panorama images taken from the real counterpart. In addition, hand tracking and virtual objects were included to both, the 3D and panorama-based environment. Therefore, an interactive VR scenario was created which is used for evaluation. In a user study, participants completed a visual search task and then filled out a questionnaire to allow comparison between the two environments. The goal of the study is to answer the following question: \emph{Is the sense of presence in VR through panorama images on par with an environment modeled in 3D?}



\section{Background and Related Work}
\label{sec:background}
Immersion is often described as the objective properties of the virtual environment that create the feeling of presence \cite{bowman2007virtual,slater1996immersion,slater1997framework}. Presence and immersion are two distinct concepts which are directly related \cite{slater2007concept} as the sense of presence is the result of immersion \cite{schubert2001experience}. Immersion, by its technical definition, is able to create a sensation of presence as described by Mestre \cite{mestre2006immersion}. Presence is the sense of an individual within an immersive environment and immersion stands for what the technology provides
from an objective point of view. As immersion is at the core of VR applications, it is an important area in research. \par 
Panorama images can be used to create immersive experiences in VR. These images are less expensive to create compared to sophisticated 3D modeled environments.
Škola et al. \cite{vskola2020virtual} combines an interactive VR experience with 360° storytelling experience in the context of cultural heritage. Using an underwater VR scenario, the authors report high levels of immersion using 360° videos. Ghida \cite{ben2020augmented} uses panorama images to lecture a history class. The authors argue that such a system could be used in the future of teaching. Metsis et al. \cite{metsis2019360} use panorama images to study and treat psychological disorders such as social anxiety. The authors use 360° videos for rapid prototyping and create therapeutic VR environments.\par 
The use of 360°/panorama images and videos is steadily increasing, but the sense of presence in a panorama-based compared to a similarly modeled 3D environment has yet to be explored.

\section{Implementation}
\label{sec:introduction}
\begin{figure*}[t]
\centering
\includegraphics[width=1.0\textwidth]{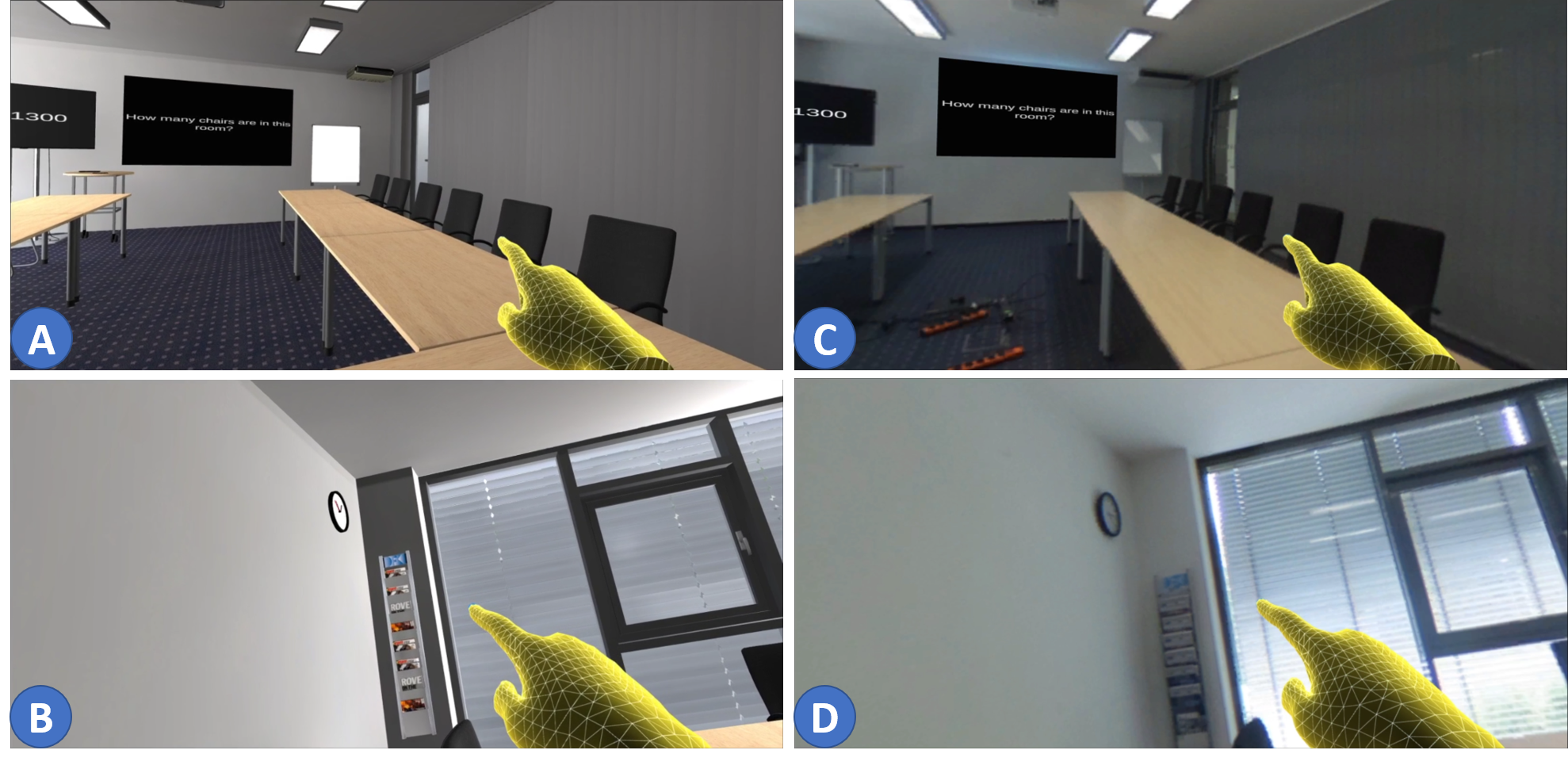}
\caption{Image (\textbf{A}) and (\textbf{B}) show the handcrafted virtual environment with 3D geometry, (\textbf{C}) and (\textbf{D}) show the virtual environment based on panorama images. The text shown in image (\textbf{C}) is superimposed onto the panorama image and shows questions to the participants. }
\label{fig:image1}
\end{figure*}

Several panorama images within a real meeting room were taken. The images were captured with a tripod on a chair at the height of a person in seating position. The images have a resolution of 5376 $x$ 2688 pixels. These images are used to create a spherical image viewer to be experienced with a VR HMD. Users are able to "sit" on each chair by switching between captured images. The selected real room has a television and projector screen. In the panorama-based virtual environment, these screens are simulated by superimposing virtual objects on the panorama images (See Figure \ref{fig:image1}\textbf{C}). Images and videos can be shown in a way that appears to the user as if it were displayed by a real TV or projector.\par
A virtual replica of this room was carefully handcrafted by an experienced 3D artist (See Figure \ref{fig:image1}\textbf{A-B}). It took about 120 working hours to recreate the room in full detail. Nearly each aspect of the real space was recreated, from the texture of the carpet to the plastic bag in the room's trash can. Furthermore, realistic lighting conditions were recreated. Behind the blinds of the windows is a high resolution HDR texture that is used to realistically illuminate the room. Figure \ref{fig:image1} shows both scenes.\par
For the experiment, interactive elements are placed in the near field of the user to answer questions shown on the projector screen. To answer these questions, the user has to enter digits on a calculator-like object with his hands.

\section{Experiment}
\subsection{Objectives}
In this experiment, the perceived realism of a handcrafted 3D environment compared to a virtual environment based on panorama images within an interactive VR scenario is investigated. In order to achieve this, selected questions of the igroup presence questionnaire\footnote{http://www.igroup.org/pq/ipq} (IPQ) \cite{schubert2001experience,regenbrecht2002real,slater1993representations} are used. Only a subset of the complete questionnaire was used since some questions were not applicable to our scenario. For example the question "I still paid attention to the real environment" is not applicable to our experiment. The list of questions is shown in section \ref{sec:results}. Subjects answered the questionnaires within the virtual environment as the work of Schwind et al. \cite{schwind2019using} suggests. 
\subsection{Participants} For this study we recruited 8 volunteers (4 Male, 4 Female). The age of the participants ranged from 31 to 60 years. All participants had no prior VR experience. \par 
\subsection{Apparatus} The experiment was conducted using an Oculus Quest 2 VR HMD connected to a gaming laptop. The resolution of the HMD is 1832 $x$ 1920 and it has a 95 degree field of view. Hand tracking and interaction with virtual objects was implemented using the Oculus SDK in the Unity game engine. \par 
\subsection{Experimental Task} The participants sit on a chair in the virtual environment. They have to perform a visual search task in which the participant is shown questions. The questions include counting the number of chairs, tables, windows, or coat hooks on the wall. There are a total of eight questions to be answered. These questions encouraged the user to fully look around in the environment. For example, to answer how many coat hooks are on the wall, the user had to turn 180° in order to solve this question. The virtual environment changes to the handcrafted or panorama-based environment after each question is answered correctly. The order of the environments was balanced. \par 
To answer the questions, a virtual object similar to a calculator is displayed to the user which can be operated hands free. After all questions are answered, the environment changes to the 3D or panorama environment respectively and the selected questions of the IPQ are answered. For this questionnaire, an object with 7 buttons is displayed that represents the answer options ranging from "fully disagree" to "fully agree". \par 
\subsection{Procedure} Participants were told to put on the VR HMD and look around to familiarize themselves with the environment. Immediately after putting on the VR HMD, the first question is visible on the projector screen. After all questions of the visual search task are completed, the selected IPQ questions are displayed. These are answered twice, once in the 3D modeled environment and once in the panorama-based environment. Each session lasted about six minutes, where participants spent three minutes in each environment.

\section{Results and Observations}
\label{sec:results}
The participants filled out a questionnaire for qualitative measurement for sense of presence in the proposed environments. Five questions from the IPQ are chosen and answered with a 7 item Likert scale:\par \noindent
\textbf{Q1: } In the computer generated world I had a sense of "being there". \par \noindent
\textbf{Q2: } Somehow I felt that the virtual world surrounded me. \par \noindent
\textbf{Q3: } How real did the virtual world seem to you? \par \noindent
\textbf{Q4: } I felt present in the virtual space. \par \noindent
\textbf{Q5: } I felt like I was just perceiving pictures. \par

\begin{figure}[ht]
\centering
\includegraphics[width=0.4\textwidth]{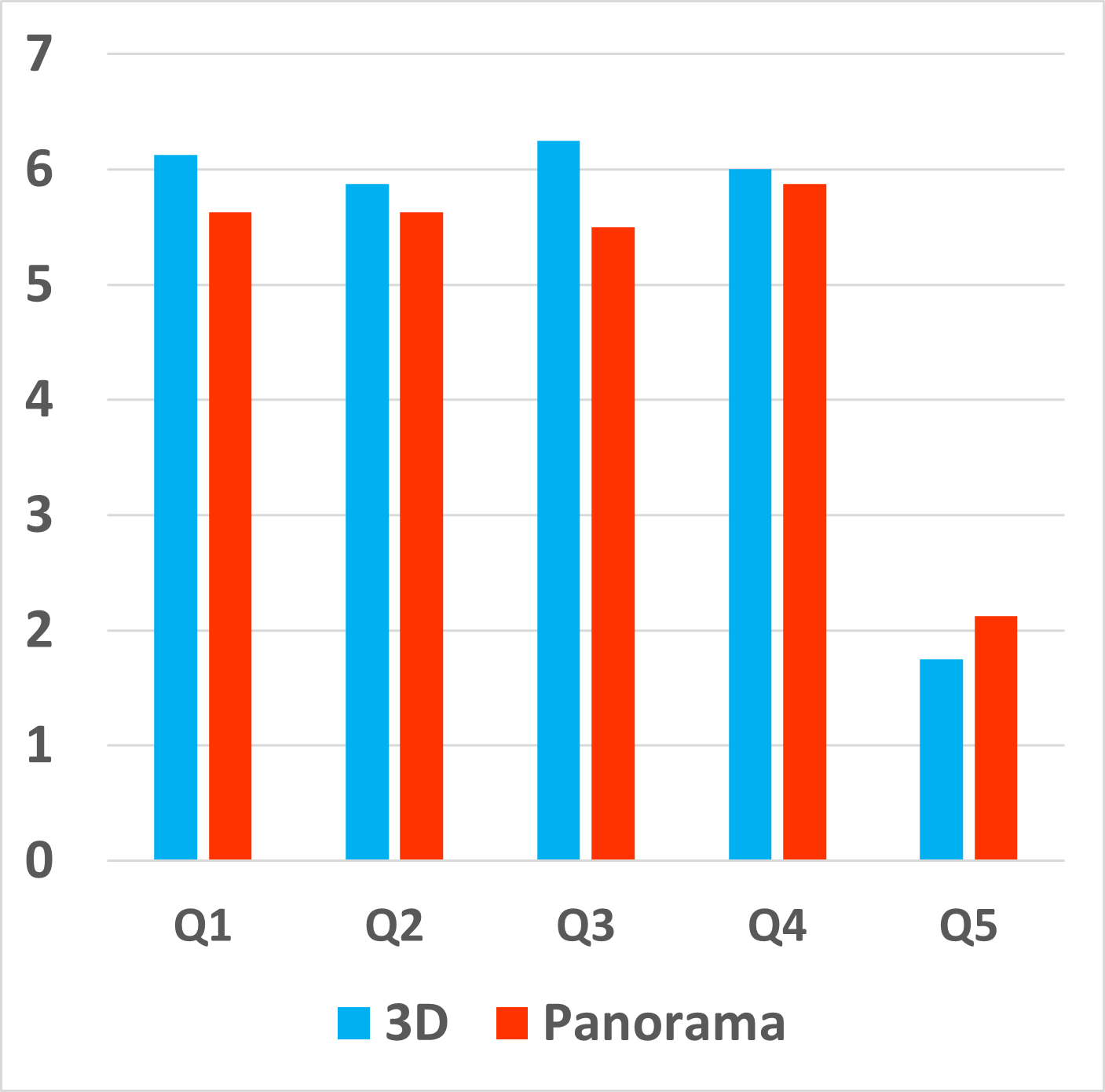}
\caption{Raw average scores from questionnaire results. A score of 1 means fully disagree and 7 means fully agree. }
\label{fig:image2}
\end{figure}

The average score for each question is shown in Figure \ref{fig:image2}. Interestingly, some participants did not notice that they observed an image instead of the 3D environment. One participant asked why the questions repeated, not knowing that the environment had changed. Some participants described the panorama based environment as pixelated compared to the handcrafted environment. Additionally, some users felt elevated in the panorama based environment.

Statistical tests were performed for each question to find significant differences between the two groups 3D environment and panorama. Levene's test assured the homogeneity of the input data for each question with p > 0.05 and therefore the data was analysed using one-way ANOVA. No significant differences between the group scores were found. The obtained p values are the following: \textbf{Q1:} $F(1,14) = 2.13$,  $p = 0.17$; \textbf{Q2:} $F(1,14) = 0.07$,  $p = 0.78$; \textbf{Q3:} $F(1,14) = 1.17$,  $p = 0.29$; \textbf{Q4:} $F(1,14) = 0.29$,  $p = 0.59$; \textbf{Q5:} $F(1,14) = 3.26$,  $p = 0.09$. \par 

\subsection{Answering the Research Question}
\emph{Is the sense of presence in VR through panorama images on par with an environment modeled in 3D?} \par 

To answer this question, the user responses as well as the questionnaire results are used. Using the five questions mentioned in section \ref{sec:results}, we found no significant different scores between the 3D modeled and the panorama-based environment. One participant did not notice that the environment had changed at all. This data and the observations lead to the conclusion that a panorama-based environment could successfully substitute a handcrafted 3D environment. However, further studies with more subjects should be conducted in order to make a conclusive assessment. \par
\subsection{Limitations}
The interactions in a panorama-based environment are limited but virtual objects can be superimposed to enable interaction. In this work, users had two screens and virtual objects with buttons (to answer the questions) as interaction possibilities. For the chosen scenario, panorama images can be a successful substitute, but further investigation for different and more complex scenarios is necessary to draw more comprehensive conclusions. \par 
Furthermore, panorama images should not be used in VR scenarios where the user is allowed to freely move around but rather in situations where he is allowed to teleport to specified points in the virtual environment. However, panorama images should be considered as a low-cost alternative in cases were users are not allowed to freely move around. For example, if scenarios involve roles such as observers or referees. \par
In future work, images with increased resolution should be used as some participants perceived the panorama environment as blurry which was caused by the comparatively low resolution of the images. 

\section{Conclusion and Future Work}
This work suggests that panorama images can be a viable alternative to handcrafted virtual environments. Depending on the context, panorama images can be an affordable and effective substitute for carefully crafted virtual environments. According to the study, users did not experience a significant difference in the sense of presence within the proposed environments. Although the chosen experiment is limited to a visual search task, there are many scenarios to which the results of this work are applicable. While the findings of this study are a starting point, a more complex study with a larger number of subjects will be conducted in the future. Furthermore, different scenarios should be evaluated. This work aims to encourage researchers to further investigate in this area. \par

\begin{acks}
This work was partially funded by the German Federal Ministry of Research and Education (BMBF) in the context of Offene Digitalisierungsallianz Pfalz under grant 03IHS075B and the EU Research and Innovation programme Horizon 2020 (project INFINITY) under the grant agreement ID: 883293.
\end{acks}

\bibliographystyle{ACM-Reference-Format}
\bibliography{sample-base}


\end{document}